\documentclass{llncs}
\usepackage{llncsdoc}
\usepackage{epsfig}
\begin{document}
\title{
Symbolic-numerical Algorithm for Generating\\  Cluster
Eigenfunctions:
  Identical Particles\\ with Pair Oscillator Interactions
       }
\titlerunning{Symbolic-numerical  Algorithm }
\author{
Alexander Gusev\inst{1},
Sergue Vinitsky\inst{1},
Ochbadrakh Chuluunbaatar\inst{1},\\
Vitaly Rostovtsev\inst{1},
Luong Le Hai\inst{1,2},
Vladimir Derbov\inst{3},\\
Andrzej G\'o\'zd\'z\inst{4},
Evgenii Klimov\inst{5}
}

\authorrunning{Alexander Gusev et al.}
\institute{
Joint Institute
for Nuclear Research, Dubna, Moscow Region, Russia, e-mail:
\email{gooseff@jinr.ru}
\and
Belgorod State University, Belgorod, Russia
\and
Saratov State University, Saratov, Russia
\and
{Department of Mathematical Physics, Institute of Physics, \\
University of Maria Curie--Sk\l{}odowska, Lublin, Poland}
\and Tver State University, Tver, Russia
}

\tocauthor{}
\maketitle
\begin{abstract}
The quantum model of a cluster, consisting of $A$ identical particles, coupled by
the internal pair interactions and affected by the external field of a target,
is considered. A symbolic-numerical algorithm for
generating
 $A\!\!-\!\!1$-dimensional  oscillator eigenfunctions, symmetric or antisymmetric with
 respect to permutations of $A$ identical particles in the new symmetrized coordinates, is formulated
and implemented using the  MAPLE computer algebra system.
Examples of generating the symmetrized coordinate representation for
$A\!\!-\!\!1$ dimensional oscillator functions
in one-dimensional Euclidean space are analyzed.
The approach is aimed at solving the problem of tunnelling the clusters,
consisting of several identical particles, through repulsive potential barriers
of a target.
 \footnote{The talk presented
at the 15th  International Workshop
''Computer Algebra in Scientific Computing 2013'', Berlin,
Germany, September 9-13, 2013}.
\end{abstract}

\section{
Introduction } Quantum harmonic oscillator wave functions have a
lot of applications in modern physics, particularly, as a basis
for constructing the wave functions of a quantum system,
consisting of $A$ identical particles, totally symmetric or
antisymmetric with respect to permutations of coordinates of the
particles~\cite {Moshinsky96}. Various special methods,
algorithms, and programs (see, e.g.,
\cite{Moshinsky96,KramerMoshinsky,Moshinsky2,Moshinsky2a,Leblond1966,Smirnov69,Novoselsky1989,WildermuthTang1977,Gintas2001})
were used to construct the desired solutions in the form of linear
combinations of the eigenfunctions of an $A-1$-dimensional
harmonic oscillator that are totally symmetric (or antisymmetric)
with respect to the coordinate permutations. However, the
implementation of this procedure in closed analytical form is
still an open problem \cite{Wang}.

A promising approach to the construction of oscillator basis
functions for four identical particles was proposed in
\cite{KramerMoshinsky,Moshinsky2,Moshinsky2a}. It was demonstrated
that a clear algorithm for generating symmetric (S) and
antisymmetric (A) states can be obtained using the symmetrized
coordinates instead of the conventional Jacobi coordinates.
However, until now this approach was not generalized for a quantum
system comprising an arbitrary number $A$ of identical particles.

We intend to develop this approach  in order to describe the
tunnelling of clusters, consisting of several coupled identical
particles, through repulsive potential barriers of a target.
Previously this problem was solved only for a pair of coupled
particles~\cite{P00P,casc11}. The developed approach will be also
applicable to the microscopic study of tetrahedral- and
octahedral-symmetric nuclei~\cite{DGZ2011} that can be considered
in the basis of seven-dimensional harmonic oscillator
eigenfunctions~\cite{Artur2012}. {\it The aim of this paper is to
present a convenient formulation of the problem stated above and
the calculation methods, algorithms, and programs for solving it.}

In this paper, we consider the quantum model of a cluster,
consisting of $A$ identical particles with the internal pair
interactions, under the influence of the external field of a
target. We assume that the spin part of the wave function is
known, so that only the spatial part of the wave function is to be
considered, which can be either symmetric or antisymmetric with
respect to a permutation of $A$ identical
particles~\cite{VAFock1930,Hamermesh,Hidaka}. The initial problem
is reduced to the problem for a composite system whose internal
degrees of freedom describe an $(A-1)\times d$-dimensional
oscillator, and the external degrees of freedom describe the
center-of-mass motion of $A$ particles in the $d$-dimensional
Euclidean space. For simplicity, we restrict our consideration to
the so-called $s$-wave approximation \cite{P00P} corresponding to
one-dimensional Euclidean space ($d=1$). It is shown that the
reduction is provided by using appropriately chosen symmetrized
coordinates rather than the conventional Jacoby coordinates.

The main goal of introducing the symmetrized coordinates is to
provide the invariance of the Hamiltonian with respect to permutations
of $A$ identical particles. This allows construction not only of
basis functions, symmetric or antisymmetric under permutations of
$A-1$ relative coordinates, but also of basis functions, symmetric
(S) or antisymmetric (A) under permutations of $A$ Cartesian
coordinates of the initial particles.
We refer the expansion of the solution in the basis of such type as the
Symmetrized Coordinate Representation (SCR).

The paper is organized as follows. In Section 2, we present the
statement of the problem in the conventional Jacobi and the
symmetrized coordinates. In Section 3, we introduce the SCR of the
solution of the considered problem and describe the appropriate
algorithm implemented using the MAPLE computer algebra system. In
Section 4, we analyze some examples of generating the symmetrized
coordinate representation for $A-1$-dimensional oscillator
functions in one-dimensional Euclidean space. In Conclusion, we
summarize the results and discuss briefly the prospects of
application of the developed approach.

\section{Problem Statement}

Consider the system of $A$ identical quantum particles with the
mass $m$ and the set of Cartesian coordinates $x_i\in {\bf R}^{d}$
in the $d$-dimensional Euclidean space, considered as the vector
$\tilde {\bf x}=(\tilde x_1,...,\tilde x_A)\in {\bf R}^{A\times
d}$ in  the $A\times d$-dimensional configuration space. The
particles are coupled by the pair potential $\tilde
V^{pair}(\tilde x_{ij})$  depending on the relative positions,
$\tilde x_{ij}=\tilde x_i-\tilde x_j$, similar to that of a
harmonic oscillator $\tilde V^{hosc}(\tilde
x_{ij})=\frac{m\omega^2}{2}(\tilde x_{ij})^2$ with the frequency
$\omega$. The whole system is subject to the influence of the
potentials $\tilde V(\tilde x_i)$ describing the external field of
a target. The system is described by the Schr\"odinger equation
\begin{eqnarray*}\left[\!-\!\frac{\hbar^2}{2m}\sum_{i=1}^A\frac{\partial^2}{\partial \tilde x_i^2}
\!+\!\sum_{i,j=1; i<j}^A \tilde V^{pair}(\tilde x_{ij}) 
\!+\!\sum_{i=1}^A\tilde V(\tilde x_i)
\!-\!\tilde E\right]
\tilde\Psi(\tilde {\bf x})\!=\!0,
\end{eqnarray*}
where $\tilde E$ is the total energy of the system of $A$ particles and
 $\tilde P^2={2m\tilde E}/{\hbar^2}$, $\tilde P$ is the total momentum of the system, and $\hbar$ is Planck constant.
Using the oscillator units $x_{osc}=\sqrt{\hbar/(m\omega\sqrt{A})}$, $p_{osc}=\sqrt{(m\omega\sqrt{A})/\hbar}=x_{osc}^{-1}$,
and $E_{osc}=\hbar\omega\sqrt{A}/2$ to introduce the  dimensionless coordinates $x_i=\tilde x_i/x_{osc}$, $x_{ij}=\tilde x_{ij}/x_{osc}=x_i-x_j$, $E=\tilde E/E_{osc}=P^2$, $P=\tilde P/p_{osc}=\tilde Px_{osc}$,
$V^{pair}(x_{ij})=\tilde V^{pair}(x_{ij} x_{osc})/E_{osc}$,
$V^{hosc}(x_{ij})=\tilde V^{hosc}(x_{ij} x_{osc})/E_{osc}=\frac{1}{A}(x_{ij})^2$
and $V(x_i)=\tilde V(x_i x_{osc})/E_{osc}$,
one can rewrite the above equation in the form
\begin{eqnarray}\left[\!-\!\sum_{i=1}^A\frac{\partial^2}{\partial x_i^2}
\!+\!\!\sum_{i,j=1; i<j}^A \frac{1}{A}(x_{ij})^2
\!+\!\!\sum_{i,j=1; i<j}^A\!U^{pair}(x_{ij})
\!+\!\!\sum_{i=1}^A V(x_i)\!-\! E\right]
\Psi( {\bf x})\!=\!0,
\label{mo1}
\end{eqnarray}
where $U^{pair}(x_{ij})=V^{pair}(x_{ij})-V^{hosc}(x_{ij})$, i.e., if  $V^{pair}(x_{ij})=V^{hosc}(x_{ij})$,
then $U^{pair}(x_{ij})=0$.

Our goal is to find the solutions
$\Psi(x_1,...,x_A)$
of Eq. (\ref{mo1}), totally symmetric (or antisymmetric)  with respect to the permutations of $A$ particles that belong to the permutation group $S_n$~\cite{Hamermesh}. The permutation of particles is nothing but a permutation of the Cartesian coordinates $x_i \leftrightarrow x_j$,   $i,j=1,...,A$.
First we introduce the Jacobi coordinates, $y= J x$, following one of the possible definitions:
\begin{eqnarray}
y_0=\frac{1}{\sqrt{A}}\left(\sum_{t=1}^{A} x_t \right),~~ y_s=\frac{1}{\sqrt{s(s+1)}}\left(\sum_{t=1}^{s} x_t-sx_{s+1}\right),~~s=1,\!...\!,A-1.\label{Ja}
\end{eqnarray}
In the matrix form Eqs. (\ref{Ja}) read as
{\small \begin{eqnarray*}
\left(\begin{array}{c} y_{0}\\y_1\\ y_2\\ y_{3}\\ \vdots\\ y_{A-1}\end{array}\right)
=J
\left(\begin{array}{c} x_1\\ x_2\\ x_{3}\\ \vdots\\ x_{A-1}\\ x_{A} \end{array}\right)
,\quad J=
\left(\begin{array}{cccccc}
1/\sqrt{A}&1/\sqrt{A}&1/\sqrt{A}&1/\sqrt{A}&\cdots&1/\sqrt{A}\\
1/\sqrt{2}&-1/\sqrt{2}&0&0&\cdots&0\\
1/\sqrt{6}&1/\sqrt{6}&-2/\sqrt{6}&0&\cdots&0\\
1/\sqrt{12}&1/\sqrt{12}&1/\sqrt{12}&-3/\sqrt{12}&\cdots&0\\
\vdots&\vdots&    \vdots&\vdots&\ddots&\vdots\\
\frac1{\sqrt{A^2-A}}&\frac1{\sqrt{A^2-A}}&\frac1{\sqrt{A^2-A}}&\frac1{\sqrt{A^2-A}}&\cdots&-\frac{A-1}{\sqrt{A^2-A}}\\
\end{array}\right),
\end{eqnarray*}}
The inverse coordinate transformation $x= J^{-1}y$ is implemented
using the transposed matrix $J^{-1}=J^T$, i.e., $J$ is an
orthogonal matrix with pairs of complex conjugate eigenvalues, the
absolute values of which are equal to one. The Jacobi coordinates
have the property $\sum_{i=0}^{A-1}(y_i\cdot
y_i)=\sum_{i=1}^A(x_i\cdot x_i)=r^2$. Therefore,
\begin{eqnarray*}
\sum_{i,j=1}^A(x_{ij})^2=2A\sum_{i=0}^{A-1}(y_i)^2-2(\sum_{i=1}^Ax_i)^2=2A\sum_{i=1}^{A-1}(y_i)^2,
\end{eqnarray*}
so that Eq. (\ref{mo1}) takes the form
\begin{eqnarray*}\nonumber
\left[-\frac{\partial^2}{\partial y_0^2}+\sum_{i=1}^{A-1}\left(-\frac{\partial^2}{\partial y_i^2}
+ (y_i)^2\right)
+U(y_0,...,y_{A-1})- E\right]\Psi(y_0,...,y_{A-1}) =0,
\\
U(y_0,...,y_{A-1})=\sum_{i,j=1; i<j}^A  U^{pair}(x_{ij}(y_1,...,y_{A-1}))
+\sum_{i=1}^A V(x_i(y_0,...,y_{A-1})),
\end{eqnarray*}
which, as follows from Eq. (\ref{Ja}), is \textit{not invariant} with respect to permutations
$y_i\leftrightarrow y_j$ at $i,j=1,...,A-1$.

\subsection*{
Symmetrized Coordinates } The  transformation from the Cartesian
coordinates to one of the possible choices of the symmetrized ones
$\xi_i$ has the form, $\xi=Cx$ and $x=C\xi$:
\begin{eqnarray*}
\xi_0=\frac{1}{\sqrt{A}}\left(\sum_{t=1}^{A} x_t \right),\quad
\xi_s=\frac{1}{\sqrt{A}}\left(x_1+ \sum_{t=2}^{A} a_0x_t+\sqrt{A}x_{s+1}\right),~~ s=1,...,A-1,
\\
x_1=\frac{1}{\sqrt{A}}\left(\sum_{t=0}^{A-1} \xi_t \right),\quad
x_s=\frac{1}{\sqrt{A}}\left(\xi_0+ \sum_{t=1}^{A-1} a_0\xi_t+\sqrt{A}\xi_{s-1}\right),\quad s=2,...,A,
\end{eqnarray*}
or, in the matrix form,
\begin{eqnarray}
\left(\begin{array}{c} \xi_0\\ \xi_1\\ \xi_{2}\\ \vdots\\ \xi_{A-2}\\ \xi_{A-1} \end{array}\right)
=C\left(\begin{array}{c} x_1\\x_2\\ x_{3}\\ \vdots\\ x_{A-1}\\ x_{A} \end{array}\right),
~~ C=\frac{1}{\sqrt{A}}
\left(\begin{array}{cccccccc}
1&1&1&1&\cdots&1&1\\
1&    a_1&    a_0&a_0&\cdots&a_{0}&a_{0}\\
1&    a_{0}&    a_1&    a_0&\cdots&a_{0}&a_{0}\\
1&    a_{0}&    a_{0}&    a_1&\cdots&a_{0}&a_{0}\\
\vdots&\vdots&    \vdots&\vdots&\ddots&\vdots&\vdots\\
1&a_{0}&    a_0&a_0&\cdots&a_{1}&a_{0}\\
1&a_{0}&    a_0&a_0&\cdots&a_{0}&a_{1}\\
\end{array}\right), \label{a}
\end{eqnarray}
where $a_0=  {1}/({1-\sqrt{A}})<0$, $a_1=a_0+\sqrt{A}$. {The
inverse coordinate transformation is performed using the same
matrix $C^{-1}=C$, $C^{2}=I$, i. e.,
 $C=C^T$ is a symmetric orthogonal  matrix
with the eigenvalues $\lambda_1=-1$, $\lambda_2=1$, ...,
$\lambda_A=1$ and det$C=-1$.} For $A=2$, the symmetrized variables
(\ref{a}) are within normalization factors similar to the
symmetrized Jacobi coordinates (\ref{Ja}) considered in
\cite{Gintas2001}, while at $A=4$ they correspond to another
choice of symmetrized coordinates $(\ddot x_4,\ddot x_1,\ddot
x_2,\ddot x_3)^T=C(x_4,x_1,x_2,x_3)^T$ considered in
\cite{KramerMoshinsky,Moshinsky2,Moshinsky2a} and mentioned
earlier in ~\cite{Hirschfelder1959,Leblond1966}. We could not find
a general definition of symmetrized coordinates for A-identical
particles like (\ref{a}) in the available  literature, so we
believe that in the present paper it is introduced for the first
time. With the relations $a_1-a_0=\sqrt{A}$, $a_0-1=a_0\sqrt{A}$
taken into into account, the relative coordinates $x_{ij}\equiv
x_i-x_j$ of a pair of particles $i$ and $j$ are expressed in terms
of the internal $A-1$ symmetrized coordinates only:
\begin{eqnarray}\nonumber
x_{ij}\equiv x_i-x_j=\xi_{i-1}-\xi_{j-1}\equiv\xi_{i-1,j-1},\\
\quad  x_{i1}\equiv x_i-x_1=\xi_{i-1}+a_0\sum_{i'=1}^{A-1}\xi_{i'},\quad i,j=2,...,A.
\label{via}\end{eqnarray}
So, if only the absolute values of $x_{ij}$ are to be considered, then there are $(A-1)(A-2)/2$
 old relative coordinates transformed into new relative ones and $A-1$ old relative coordinates
 expressed in terms of $A-1$ internal symmetrized coordinates.
These important relations essentially simplify the procedures of
symmetrization (or antisymmetrization) of the oscillator basis
functions and the calculations of the corresponding pair-interaction
integrals  $V^{pair}(x_{ij})$.
The symmetrized coordinates are related to the Jacobi ones as
$y = B \xi$, $B=J C$:
\begin{eqnarray}\label{b}\left(\begin{array}{c} y_0\\ y_1\\ y_{2}\\ \vdots\\ y_{A-2}\\ y_{A-1}\end{array}\right)
=
B\left(\begin{array}{c} \xi_0\\ \xi_1\\ \xi_{2}\\ \vdots\\ \xi_{A-2}\\ \xi_{A-1} \end{array}\right)
,~~
B 
=\left(\begin{array}{cccccccc}
1&0&0&0&0&\cdots&0&0\\
0&b_1^0&b_1^-&b_1^-&b_1^-&\cdots&b_1^-&b_1^-\\
0&b_2^+&b_2^0&b_2^-&b_2^-&\cdots&b_2^-&b_2^-\\
0&b_3^+&b_3^+&b_3^0&b_3^-&\cdots&b_3^-&b_3^-\\
0&b_4^+&b_4^+&b_4^+&b_4^0&\cdots&b_4^-&b_4^-\\
\vdots&\vdots&    \vdots&\vdots&\vdots&\ddots&\vdots&\vdots\\
0&b_{A-1}^+&b_{A-1}^+&b_{A-1}^+&b_{A-1}^+&\cdots&b_{A-1}^+&b_{A-1}^0\\
\end{array}\right),
\end{eqnarray}
where $b_{s}^+=1/((\sqrt{A}-1)\sqrt{s(s+1)})$,
$b_{s}^-=\sqrt{A}/((\sqrt{A}-1)\sqrt{s(s+1)})$, and
$b_{s}^0=(1+s-s\sqrt{A})/((\sqrt{A}-1)\sqrt{s(s+1)})$. One can see
that for the center of mass the symmetrized and Jacobi coordinates
are equal, $y_0=\xi_0$, while the relative coordinates are related
via the $(A-1)\times(A-1)$ matrix $M$ with the elements
$M_{ij}=B_{i+1,j+1}$ and det$M=(-1)^{A\times d}$, i.e.,  the
matrix, obtained by cancelling the first row and the first column.
The inverse transformation $\xi = B^{-1} y $ is given by the
matrix $B^{-1}=(J C)^{-1}=CJ^{T}=B^T$, i.e., $B$ is also an
orthogonal matrix.

In the symmetrized coordinates Eq. (\ref{mo1})
takes the form
\begin{eqnarray}\left[-\frac{\partial^2}{\partial \xi_0^2}
+\sum_{i=1}^{A-1}\left(-\frac{\partial^2}{\partial \xi_i^2}
+ (\xi_i)^2\right)
+U(\xi_0,...,\xi_{A-1})
- E\right] \Psi(\xi_0,...,\xi_{A-1}) =0,\label{mo5}
\\ \nonumber
U(\xi_0,...,\xi_{A-1})=\sum_{i,j=1; i<j}^A  U^{pair}(x_{ij}(\xi_1,...,\xi_{A-1}))
+\sum_{i=1}^A V(x_i(\xi_0,...,\xi_{A-1})),
\end{eqnarray}
which is \textit{invariant} under permutations $\xi_i\leftrightarrow \xi_j$ at $i,j=1,...,A-1$, as follows from Eq. (\ref{a}), i.e., the \textit{invariance} of Eq. (\ref{mo1}) under permutations $x_i\leftrightarrow x_j$ at $i,j=1,...,A$ survives.
\begin{table}[t]
\caption{The first few eigenvalues $E_j^S$ and the oscillator S-eigenfunctions
(\ref{fbf})
at $E_j^S-E_1^S\le 10$, $E_1^S=A-1$. We use the notations
$|[i_1,i_2,...,i_{A-1}]\rangle
\equiv\Phi^s_{[i_1,i_2,...,i_{A-1}]}(\xi_1,...,\xi_{A-1})$ from
Eqs. (\ref{nbeta}) and (\ref{ant0}), i.e., $[i_1,i_2,...,i_{A-1}]$
assumes the summation over  permutations of
$[i_1,i_2,...,i_{A-1}]$ in the layer
$2\sum_{k=1}^{A-1}i_k+A-1=E_i^{s(a)}$. }\label{18}
\begin{center}
\begin{tabular}{||c|c||c|c||c|c||c||} \hline
\multicolumn{2}{||c||}{A=2}    & \multicolumn{2}{|c||}{A=3}  & \multicolumn{2}{|c||}{A=4}     & $E_j^S-E_1^S$\\ \hline
j&$\Phi_j^S(\xi_1)$&j&$\Phi_j^S(\xi_1,\xi_{2})$&j&$\Phi_j^S(\xi_1,\xi_{2},\xi_{3})$&        \\  \hline
1&$|[0]\rangle $&1  & $|[0, 0]\rangle $                   & 1  &$|[0, 0, 0]\rangle $ &0   \\ \hline
2&$|[2]\rangle $&2  & $|[0, 2]\rangle $                   & 2  &$|[0, 0, 2]\rangle $ &4   \\ \hline
 &     &3  &$\frac{1}{2}|[0, 3]\rangle -\frac{\sqrt{3}}{2}|[1, 2]\rangle $
                                        &  3 &$|[1, 1, 1]\rangle $ &6   \\ \hline
  3&$|[4]\rangle $&4  & $\frac{\sqrt{3}}{2}|[0, 4]\rangle +\frac{1}{2}|[2, 2]\rangle $ & 4  &$|[0, 0, 4]\rangle $     &8   \\ \cline{5-7}
   &$ $&   &   & 5  &$|[0, 2, 2]\rangle $  &8   \\ \hline
   &       &5  &
   $\frac{\sqrt{5}}{4}|[0,5]\rangle-\frac{ {3}}{4}|[1,4]\rangle
   -\frac{\sqrt{2}}{4}|[2,3]\rangle $
    & 6 & $|[1,1,3]\rangle $  &10  \\
 \hline
\end{tabular}\end{center}
\end{table}
\begin{table}[t]
\caption{The first few eigenvalues $E_j^A$ and the oscillator A-eigenfunctions
(\ref{fbf})
at $E_j^A-E_1^A\le 10$, $E_1^A=A^2-1$. We use the notations
$|[i_1,i_2,...,i_{A-1}]\rangle
\equiv\Phi^a_{[i_1,i_2,...,i_{A-1}]}(\xi_1,...,\xi_{A-1})$ from
Eq. (\ref{ant1}), i.e., $[i_1,i_2,...,i_{A-1}]$ assumes the
summation over the multiset permutations of
$[i_1,i_2,...,i_{A-1}]$ in the layer
$2\sum_{k=1}^{A-1}i_k+A-1=E_i^{s(a)}$.}\label{18f}
\begin{center}
\begin{tabular}{||c|c||c|c||c|c||c|c||c|c||c||} \hline
\multicolumn{2}{||c||}{$A=2$, $E_1^A=3$}    & \multicolumn{2}{|c||}{$A=3$, $E_1^A=8$}  & \multicolumn{2}{|c||}{$A=4$, $E_1^A=15$}
& $E_j^A-E_1^A$\\ \hline
j&$\Phi_j^A(\xi_1)$&j&$\Phi_j^A(\xi_1,\xi_{2})$&j&$\Phi_j^A(\xi_1,\xi_{2},\xi_{3})$ &       \\  \hline
1&$|[1]\rangle $&1  & $\frac{1}{2}|[0, 3]\rangle +\frac{\sqrt{3}}{2}|[1, 2]\rangle $
& 1  &$|[0, 2, 4]\rangle $  &0
\\  \hline
2&$|[3]\rangle $&2  & $\frac{\sqrt{5}}{4}|[0, 5]\rangle \!+\!\frac{ {3}}{4}|[1, 4]\rangle \!-\!\frac{\sqrt{2}}{4}|[2, 3]\rangle $
& 2  &$|[0, 2, 6]\rangle $  &4   \\ \hline
 &     &3  &$\frac{1}{4}|[0, 6]\rangle -\frac{\sqrt{15}}{4}|[2, 4]\rangle $  &
 3 &$|[1, 3, 5]\rangle $  &6   \\  \hline
3&$|[5]\rangle $&4  & $\frac{\sqrt{21}}{8}|[0, 7]\rangle +\frac{3\sqrt{3}}{8}|[1, 6]\rangle$ &
4  &$|[0, 4, 6]\rangle $  &8   \\ \cline{5-7}
   &     &   &$~~~~~-\frac{1}{8}|[2, 5]\rangle+\frac{\sqrt{5}}{8}|[3, 4]\rangle $
   & 5  &$|[0, 2, 8]\rangle $
    &8   \\ \hline
   &       &5  & $\frac{\sqrt{2}}{4}|[0, 8]\rangle -\frac{\sqrt{14}}{4}|[2, 6]\rangle $  & 6 & $|[1,3,7]\rangle $    &10  \\
 \hline
\end{tabular}\end{center}
\end{table}
\section{The SCR Algorithm: Symmetrized Coordinate Representation}\label{sec3}

For simplicity, consider the solutions of Eq. (\ref{mo5}) in the internal symmetrized coordinates $\{\xi_1,...,\xi_{A-1}\}\in {\bf R}^{A-1}$, $  x_i\in {\bf R}^1$,
 in the case of 1D Euclidean space ($d=1$). The relevant equation describes
 an $(A-1)$-dimensional oscillator with the eigenfunctions $\Phi_j(\xi_1,...,\xi_{A-1})$ and  the energy eigenvalues $E_j$:
\begin{eqnarray} \left[\sum_{i=1}^{A-1}\left(-\frac{\partial^2}{\partial \xi_i^2}\!+ \!(\xi_i)^2\right)\!-\! E_j\right] \Phi_j(\xi_1,...,\xi_{A-1}) \!=\!0,\quad E_j=2\!\sum\limits_{k=1}^{A-1}\!i_k\!+\!A\!-\!1, \label{mo2g}
\end{eqnarray}
where the numbers $i_k$, $k=1,...,A-1$ are integer, $i_k=0,1,2,3,...$.
The eigenfunctions $\Phi_j(\xi_1,...,\xi_{A-1})$ can be expressed in terms of the conventional eigenfunctions of individual 1D oscillators as
\begin{eqnarray}&&\label{nbeta}
\Phi_j(\xi_1,...,\xi_{A-1})=\!\!\!\sum\nolimits_{2\sum\limits_{k=1}^{A-1}i_k+A-1=E_j}\!\!\!\!\!\!\!
\beta_{j[i_1,i_2,...,i_{A-1}]}\bar\Phi_{[i_1,i_2,...,i_{A-1}]}(\xi_1,...,\xi_{A-1}), \\&&\nonumber \bar\Phi_{[i_1,i_2,...,i_{A-1}]}(\xi_1,...,\xi_{A-1}) =
\prod_{k=1}^{A-1}\bar\Phi_{i_k}(\xi_k),\quad
\bar\Phi_{i_k}(\xi_k)=\frac{\exp(-\xi_k^2/2)H_{i_k}(\xi_k)}{\sqrt[4]\pi\sqrt{2^{i_k}}\sqrt{{i_k}!}},
\end{eqnarray}
where $H_{i_k}(\xi_k)$ are Hermite polynomials \cite{stigun}.
Generally the energy level $E_f=2f+A-1$,
$f=\sum\nolimits_{k=1}^{A-1}i_k$, of an $(A-1)$-dimensional oscillator is known \cite{Baker} to possess the
degeneracy multiplicity $p=(A+f-2)!/f!/(A-2)!$ with respect to the conventional oscillator eigenfunctions $\bar\Phi_{[i_1,i_2,...,i_{A-1}]}(\xi_1,...,\xi_{A-1})$. This degeneracy allows further symmetrization by choosing the appropriate coefficients $\beta^{(j)}_{[i_1,i_2,...,i_{A-1}]}$.
Degeneracy multiplicity $p$ of all states  with the given energy $E_j$ defined by formula
\begin{eqnarray}&&\label{nbetab}
p=\sum\nolimits_{2\sum\nolimits_{k=1}^{A-1}i_k+A-1=E_j}N_\beta, \quad N_\beta=(A-1)!/\prod_{k=1}^{N_\upsilon}\upsilon_k!,
\end{eqnarray} 
where $N_\beta$ is the number of multiset permutations (m.p.) of $[i_1,i_2,...,i_{A-1}]$,
and  $N_\upsilon\leq A-1$ is the number of different values $i_k$ in the multiset $[i_1,i_2,...,i_{A-1}]$, and
 $\upsilon_k$ is the number of repetitions of the given value $i_k$.

\noindent{\bf Step 1. Symmetrization with respect to permutation of $A\!\!-\!\!1$ particles}\\
For the states $\Phi_j^s(\xi_1,...,\xi_{A-1})\equiv\Phi_{[i_1,i_2,...,i_{A-1}]}^s(\xi_1,...,\xi_{A-1})$,  symmetric with respect to permutation of $A-1$ particles $i=[i_1,i_2,...,i_{A-1}]$,
the coefficients $\beta_{i[i_1',i_2',...,i_{A-1}']}$ in Eq. (\ref{nbeta}) are
\begin{eqnarray}\label{ant0}
\beta_{i[i_1',i_2',...,i_{A-1}']}=\left\{ \begin{array}{ll}
\frac{1}{\sqrt{N_\beta}},& \mbox{if  $[i_1',i_2',...,i_{A-1}']$ is a m. p. of   $[i_1,i_2,...,i_{A-1}]$,}\\
0,& \mbox{otherwise}.\end{array}\right.
\end{eqnarray}

The states $\Phi_j^a(\xi_1,...,\xi_{A-1})\equiv\Phi_{[i_1,i_2,...,i_{A-1}]}^a(\xi_1,...,\xi_{A-1})$, antisymmetric with respect to permutation of $A-1$ particles
are constructed in a conventional way
\begin{eqnarray}\label{ant1}
\Phi_{j}^a(\xi_1,...,\xi_{A-1})=\frac{1}{\sqrt{(A-1)!}}
\left|\begin{array}{cccc}
\bar\Phi_{i_1}(\xi_1)&\bar\Phi_{i_2}(\xi_1)&\cdots&\bar\Phi_{i_{A-1}}(\xi_1)\\
\bar\Phi_{i_1}(\xi_2)&\bar\Phi_{i_2}(\xi_2)&\cdots&\bar\Phi_{i_{A-1}}(\xi_2)\\
\vdots&\vdots&\ddots&\vdots\\
\bar\Phi_{i_1}(\xi_{A-1})&\bar\Phi_{i_2}(\xi_{A-1})&\cdots&\bar\Phi_{i_{A-1}}(\xi_{A-1})\\
\end{array}\right|,
\end{eqnarray}
i.e., the coefficients $\beta^{(i)}_{[i_1',i_2',...,i_{A-1}']}$ in (\ref{nbeta})
are expressed as $$\beta^{(i)}_{[i_1',i_2',...,i_{A-1}']}=\varepsilon_{i_1',i_2',...,i_{A-1}'}/{\sqrt{(A-1)!}},$$
where $\varepsilon_{i_1',i_2',...,i_{A-1}'}$ is a totally antisymmetric tensor. This tensor is defined as follows:
$\varepsilon_{i_1',i_2',...,i_{A-1}'}=+1(-1)$, if $i_1',i_2',...,i_{A-1}'$ is an even (odd) permutation of the numbers
 $i_1<i_2<...<i_{A-1}$, and $\varepsilon_{i_1',i_2',...,i_{A-1}'}=0$ otherwise, i.e., when some two numbers in the set ${i_1',i_2',...,i_{A-1}'}$ are equal.
 Therefore, for antisymmetric states the numbers $i_k$ in Eq. (\ref{mo2g}) take the integer values $i_k=k-1,k,k+1,...$, $k=1,...,A-1$.

Here and below  $s$ and $a$ are used for the functions, symmetric
(antisymmetric) under permutations of $A-1$ relative coordinates,
constructed at the first step of the procedure. On the contrary,
$S$ and $A$ are used for the functions, symmetric (asymmetric)
under permutations of $A$ initial Cartesian coordinates. This is
actually the symmetry with respect to permutation of identical
particles themselves; in this sense, S and A states may be
attributed to boson- and fermion-like particles. However, we
prefer to use the S (A) notation as more rigorous.

\begin{figure}[t]
 \epsfig{file=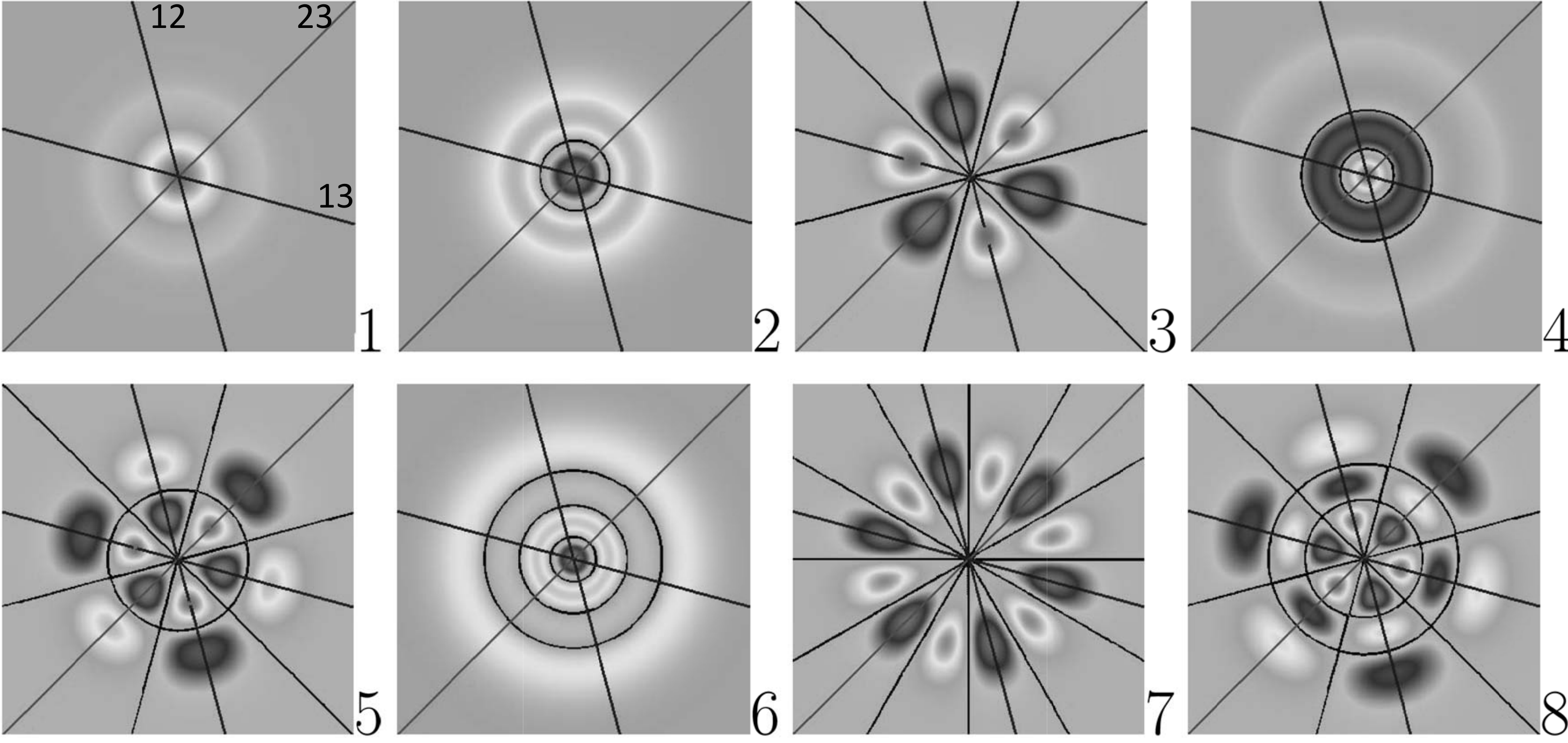,width=0.98\textwidth,angle=0 }
\caption{ Profiles of the first eight oscillator S-eigenfunctions
$\Phi^S_{[i_1,i_2]}(\xi_1,\xi_{2})$, at $A=3$ in the coordinate frame $(\xi_1,\xi_{2})$. The lines correspond to pair collision $x_2=x_3$, $x_1=x_2$ and $x_1=x_3$ of the projection $(x_1,x_2,x_3)\to(\xi_1,\xi_{2})$, marked only in the left upper panel  with `23', `12', and `13', respectively. The additional lines are nodes of the eigenfunctions $\Phi^S_{[i_1,i_2]}(\xi_1,\xi_{2})$.}
\label{table3}
\end{figure}
\begin{figure}[t]
 \epsfig{file=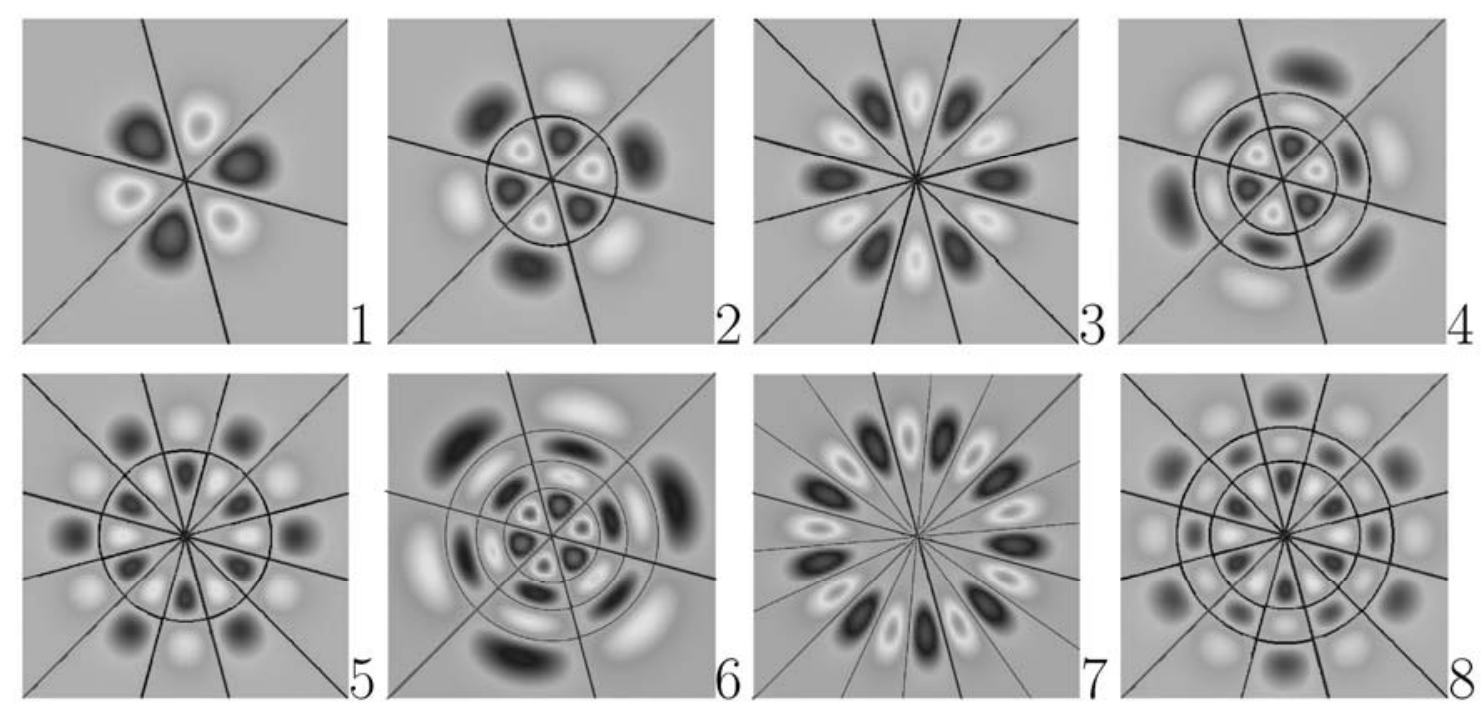,width=0.98\textwidth,angle=0 }
\caption{The same as in Fig. \ref{table3}, but for the first eight oscillator A-eigenfunctions
$\Phi^A_{[i_1,i_2]}(\xi_1,\xi_{2})$, at $A=3$. }
\label{table4}
\end{figure}

\noindent{\bf Step 2. Symmetrization with respect to permutation of $A$ particles}\\
For $A=2$, the symmetrized coordinate $\xi_1$ corresponds to the
difference $x_2-x_1$ of Cartesian coordinates, so that a function
even (odd) with respect to $\xi_1$ appears to be symmetric
(antisymmetric)  with respect to the permutation of two particles
$x_2\leftrightarrow x_1$. Hence, even (odd) eigenfunctions with
corresponding eigenvalues $E_j^s=2(2n)+1$ ($E_j^a=2(2n+1)+1$)
describe S (A) solutions.

For $A\geq 3$, the functions, symmetric (antisymmetric) with
respect to permutations of Cartesian coordinates
$x_{i+1}\leftrightarrow x_{j+1}$, $i,j=0,...,A-1$:
$$\Phi^{S(A)}(...,x_{i+1},...,x_{j+1},...)\equiv\Phi^{S(A)}(\xi_1(x_1,...,x_{A}),...,\xi_{A-1}(x_1,...,x_{A}))
$$
$$=\pm\Phi^{S(A)}(...,x_{j+1},...,x_{i+1},...)$$
become symmetric (antisymmetric)
with respect to permutations of symmetrized coordinates $\xi_i\leftrightarrow \xi_j$,  $i,j=1,...,A-1$: $$\Phi^{S(A)}(...,\xi_i,...,\xi_j,...)=\pm\Phi^{S(A)}(...,\xi_j,...,\xi_i,...),$$ as follows from Eq. (\ref{via}).
However, the converse statement is not valid,
$$\Phi^{s(a)}(...,\xi_i,...,\xi_j,...)=\pm\Phi^{s(a)}(...,\xi_j,...,\xi_i,...) ~~~~~~$$ $$\not
\Rightarrow
\Phi^{s(a)}(x_1,...,x_{i+1},...)=\pm\Phi^{s(a)}(x_{i+1},...,x_1,...),$$ because we deal with a projection map
\begin{eqnarray}\label{pmm}(\xi_1,...,\xi_{A-1})^T=\hat C(x_1,...,x_{A})^T,\end{eqnarray}
which is implemented by the $(A-1)\times(A)$ matrix $\hat C$ with the matrix elements $\hat C_{ij}=C_{i+1,j}$,
obtained from (\ref{a}) by cancelling the first row.
Hence, the functions, symmetric (antisymmetric)
with respect to permutations of symmetrized coordinates, are divided
into two types, namely, the
S (A)
solutions, symmetric (antisymmetric) with respect to permutations
$x_{1}\leftrightarrow x_{j+1}$ at $j=1,...,A-1$:
$$\Phi^{S(A)}(x_1,...,x_{j+1},...)=\pm\Phi^{S(A)}(x_{j+1},...,x_1,...)
$$
and the other s (a)
solutions,  $\Phi^{s(a)}(x_1,...,x_{i+1},...)\neq\pm\Phi^{s(a)}(x_{i+1},...,x_1,...)$, which should be eliminated.
These requirements are equivalent to only one permutation $x_1\leftrightarrow x_2$, as follows from (\ref{via}), which simplifies their practical implementation.
With these requirements taken into account in the Gram--Schmidt process, implemented in the
\textit{symbolic algorithm SCR}, we obtained the required characteristics of S and A eigenfunctions,
\begin{eqnarray}\label{fbf}
\Phi_{i}^{S(A)}(\xi_1,...,\xi_{A-1})=\!\!\!\!\!\!\!\!\!\!\sum \limits_{2\sum_{k=1}^{A-1}i_k+A-1=E_i^{s(a)}} \!\!\!\!\!\!\!\!\!\!\!\!\alpha^{S(A)}_{i{[i_1,i_2,...,i_{A-1}]}}\Phi^{s(a)}_{[i_1,i_2,...,i_{A-1}]}(\xi_1,...,\xi_{A-1}).
\end{eqnarray}

\textbf{The algorithm \textit{SCR}:}\\[-3mm]
\underline{\hspace{.995\textwidth}}\\
\textbf{Input}:\\
$A$ is the number of identical particles;\\
$i_{max}$ is defined by the maximal value of the energy $E_{i_{max}}$;\\
$(\xi_1,...,\xi_{A-1})$ and $(x_1,...,x_{A})$
are the symmetrized
and the Cartesian
coordinates; 
\\[-4mm]
\noindent \underline{\hspace{.995\textwidth}}\\
 \textbf{Output}:\\
$\Phi_{i}^{S(A)}(\xi_1,...,\xi_{A-1})$
and $\Phi_{i}^{S(A)}(x_1,...,x_{A})$ are the total symmetric (antisymmetric) functions (\ref{fbf})
in the above
coordinates connected by (\ref{pmm});\\[-4mm]
\noindent \underline{\hspace{.995\textwidth}}\\
\textbf{Local}:\\
$E_i^{s(a)}\equiv E_i^{S(A)}=2\sum_{k=1}^{A-1}i_k+A-1$ is the $(i+1)^{th}$ eigenenergy;\\
$i_{min}=0$ for the symmetric and $i_{min}=(A-1)^2$ for the antisymmetric case;\\
$\Phi_{j}\equiv\Phi^{s(a)}_{[i_1,i_2,...,i_{A-1}]}(\xi_1,...,\xi_{A-1})$ and
$\Phi_{j}\equiv\Phi^{s(a)}_{[i_1,i_2,...,i_{A-1}]}(x_1,...,x_{A})$ 
are the functions, symmetric (antisymmetric) with respect to $A-1$ Cartesian coordinates;\\
$p_{s(a)}\equiv p_{i;s(a)}$ and $p_{S(A)}\equiv p_{i;S(A)}$ are the degeneracy factors of the energy levels $E_i^{s(a)}$ and $E_i^{S(A)}$ for s(a) and S(A) functions, respectively;\\
$p_{i;min}$ ($p_{i;max}$) and $P_{i;min}$ ($P_{i;max}$) are the lowest (highest) numbers of s(a) and S(A) functions, belonging  to the energy levels $E_i^{s(a)}$ and $E_i^{S(A)}$, respectively; \\
$\{\bar\alpha_{j}\}$ and $\{\alpha^{S(A)}_{pj}\}$ are the sets of intermediate and desired coefficients;\\[-3mm]
\underline{\hspace{.995\textwidth}}\\
1.1 $j:=0$;\\
\textbf{for} $i$ from $i_{min}$ to $i_{max}$ \textbf{do};\\
1.2: $p_{i;min}:=j+1$;\\
1.3: \textbf{for each } sorted $i_1,i_2, ...,i_{A-1}$, $2\sum_{k=1}^{A-1}i_k+A-1=E_i^{s(a)}$ \textbf{do} \\
\phantom{1.1:1.11} $j:=j+1$;\\
\phantom{1.1:1.11} construction $\Phi_{j}(\xi_1,...,\xi_{A-1})=\Phi_{j}^s(\xi_1,...,\xi_{A-1})$ from (\ref{nbeta}), (\ref{ant0})\\
\phantom{1.1:1.11111}
or $\Phi_{j}(\xi_1,...,\xi_{A-1})=\Phi_{j}^a(\xi_1,...,\xi_{A-1})$ from   (\ref{ant1})\\
\phantom{1.1:1.11} $\Phi_{j}(x_1,...,x_{A})=$subs$((\xi_1,...,\xi_{A-1})\to(x_1,...,x_{A}), \Phi_{j}(\xi_1,...,\xi_{A-1}))$;\\
\phantom{1.1:}\textbf{end for}\\
1.4: $p_{i;max}:=j$;
$~~~p_{i;s(a)}=p_{i;max}-p_{i;min}+1$;\\
\textbf{end for}\\[-4mm]
\underline{\hspace{.995\textwidth}}\\
2.1.:$P_{min}=1$;\\
\textbf{for} $i$ from $i_{min}$ to $i_{max}$ \textbf{do}\\
2.2.:$P_{i;min}=P_{min}$;\\
2.3.:$\Phi(\xi_1,...,\xi_{A-1})= \sum_{j=p_{i;min}}^{p_{i;max}}\bar{\alpha}_{j}\Phi_{j}(\xi_1,...,\xi_{A-1})$;\\
\phantom{2.1.:} $\Phi(x_1,...,x_{A})= \sum_{j=p_{i;min}}^{p_{i;max}}\bar{\alpha}_{j}\Phi_{j}(x_1,...,x_{A})$;\\
2.4.: $\Phi(x_2,x_1,...,x_{A}):=$change$(x_1\leftrightarrow x_2,\Phi(x_1,x_2,...,x_{A})))$;\\
2.5.: $\Phi(x_2,x_1,...,x_{A})\mp\Phi(x_1,x_2,...,x_{A})=0,$ \\
\phantom{2.4.:} $\quad\to\quad (\bar{\alpha}_{pj}, j=p_{i;min},...,p_{i;max}, p=1,...,p_{i;S(A)})$ ;\\
2.6.:$P_{i;max}=P_{i;min}-1+p_{i;S(A)}$;\\%
2.7.: Gram--Schmidt procedure for $\Phi(\xi_1,...,\xi_{A-1})\qquad \to$\\
\phantom{2.4.:} $\Phi_{p}^{S(A)}(x_1,x_2,...,x_{A})=\sum_{j=p_{i;min}}^{p_{i;max}} \alpha^{S(A)}_{pj}\Phi_{j}(x_1,x_2,...,x_{A})$;
\\
\phantom{2.4.:} $\Phi_{p}^{S(A)}(\xi_1,...,\xi_{A-1})=\sum_{j=p_{i;min}}^{p_{i;max}} \alpha^{S(A)}_{pj}\Phi_{j}(\xi_1,...,\xi_{A-1})$,\\
\phantom{2.4.:} at $p=P_{i;min},...,P_{i;max}$;\\
\textbf{end for}\\[-4mm]
\underline{\hspace{.995\textwidth}}

\section{Examples of the  SCR Generation
}
\begin{figure}[t]
 \epsfig{file=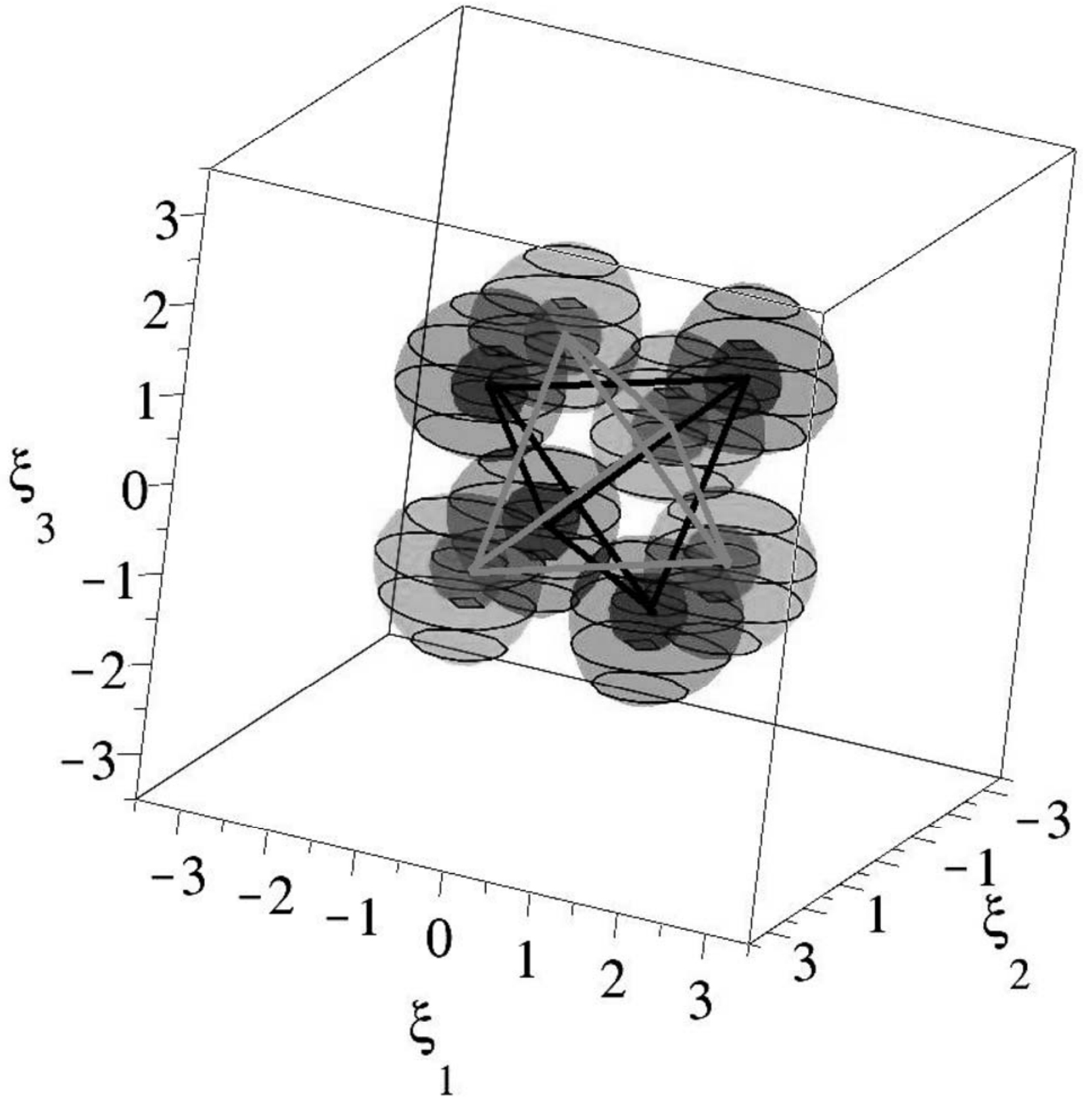,width=0.32\textwidth,height=0.32\textwidth,angle=0 }
 \epsfig{file=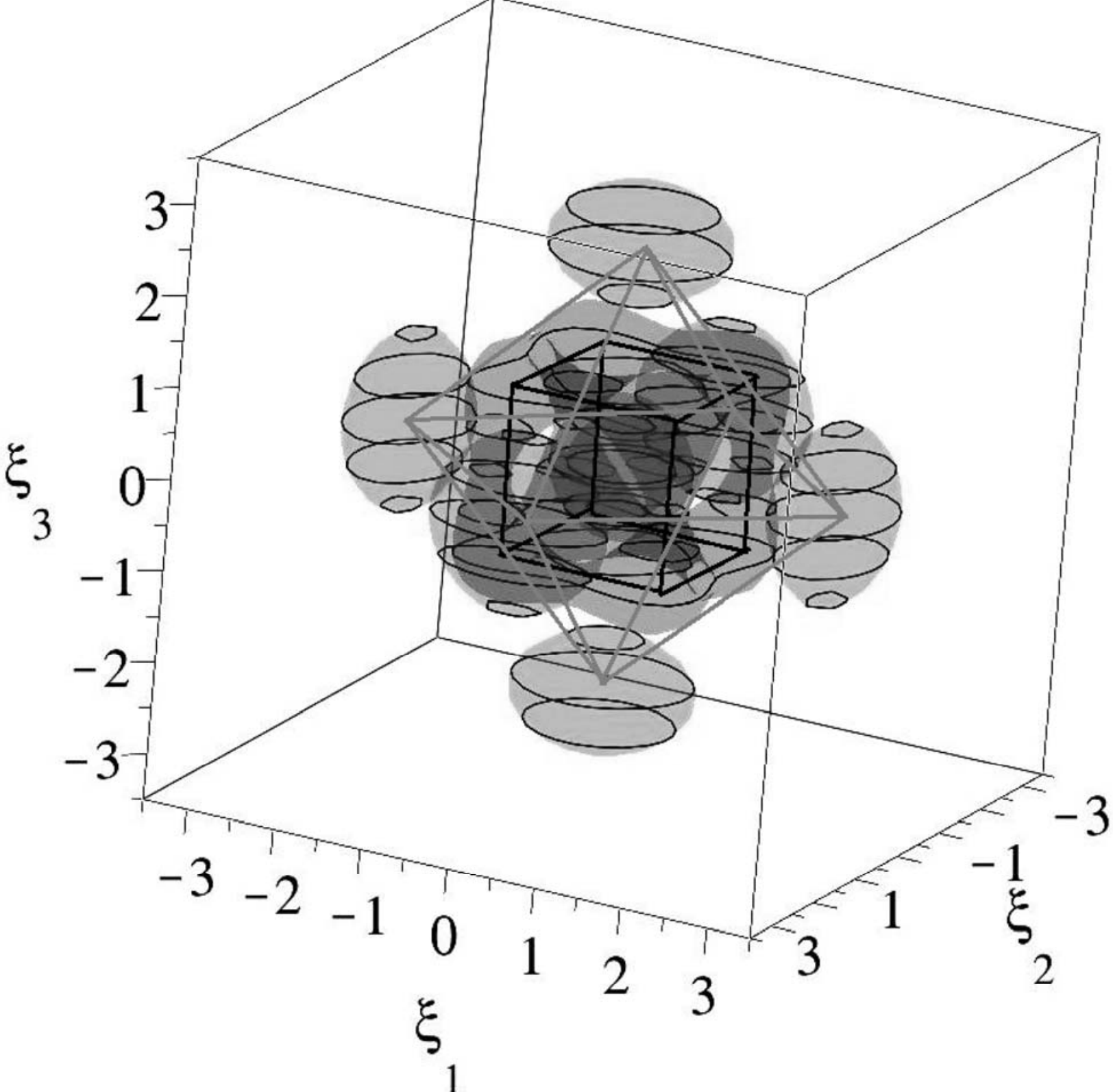,width=0.32\textwidth,height=0.32\textwidth,angle=0 }
 \epsfig{file=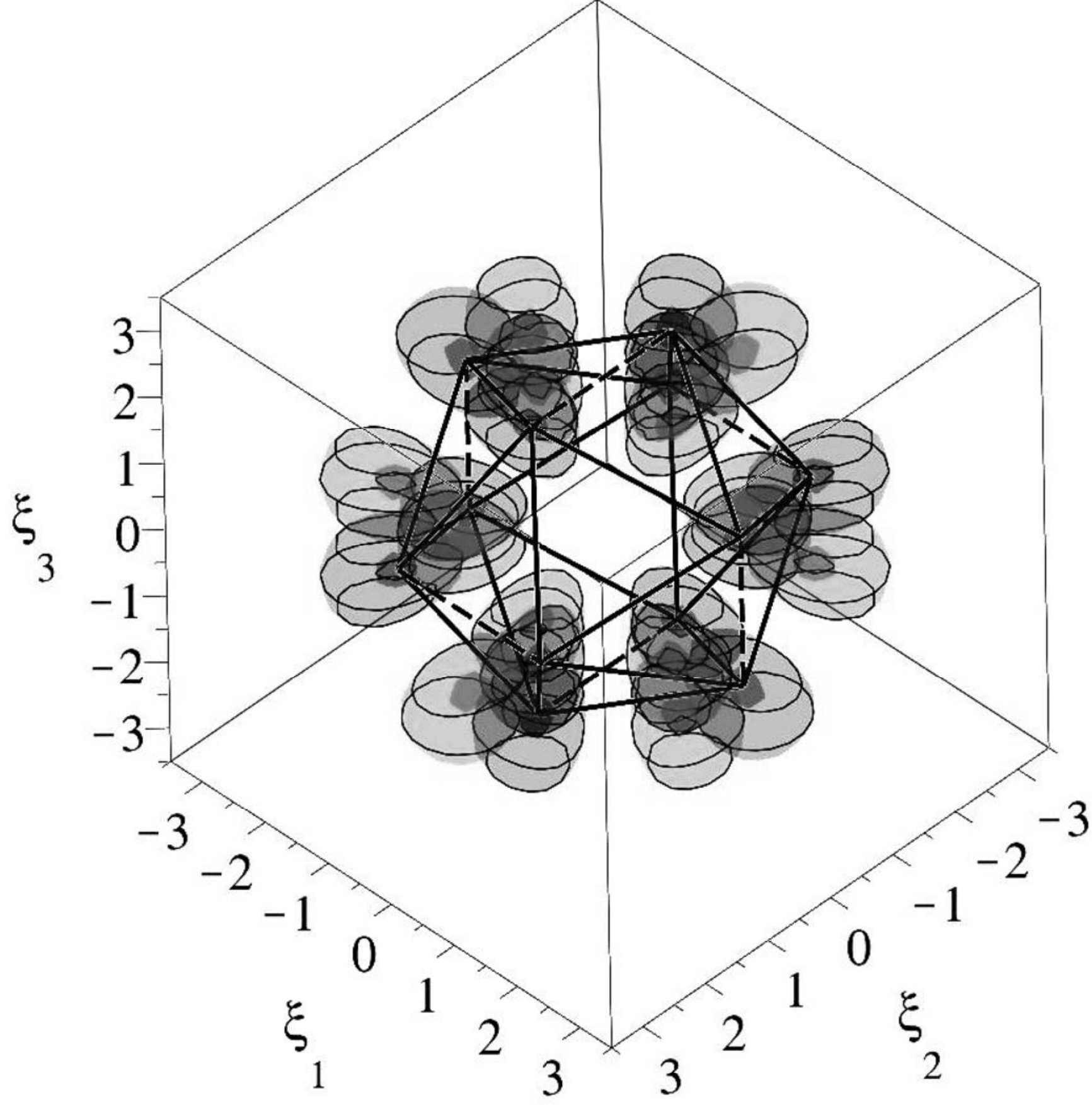,width=0.32\textwidth,height=0.32\textwidth,angle=0 }
 \epsfig{file=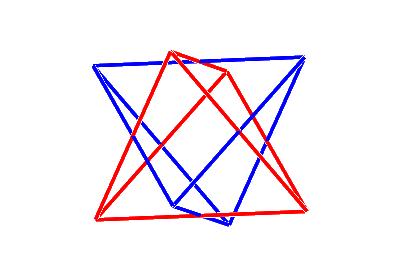,width=0.32\textwidth,angle=0 }
 \epsfig{file=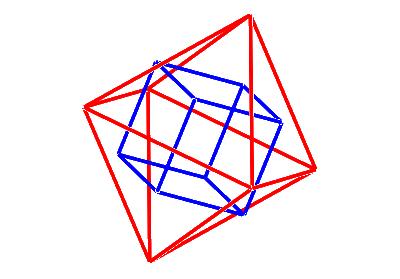,width=0.32\textwidth,angle=0 }
 \epsfig{file=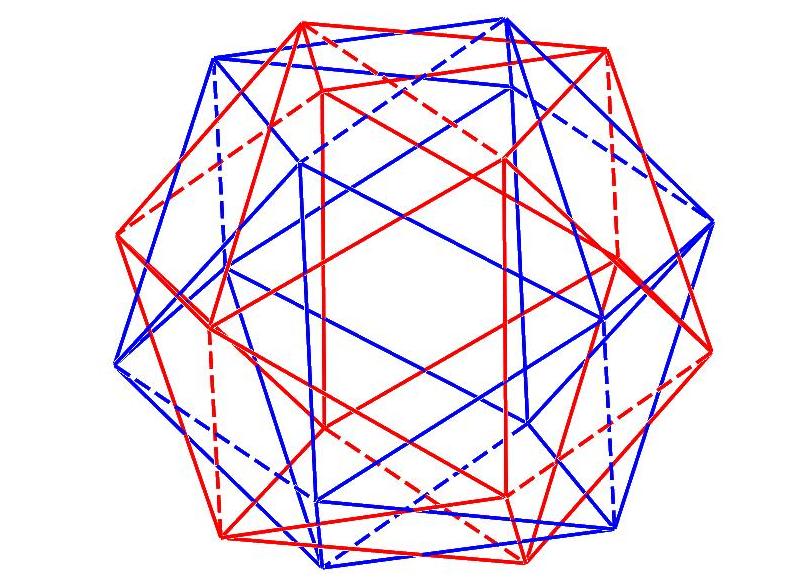,width=0.32\textwidth,angle=0 }
%
\caption{ Upper panel: Profiles of the oscillator S-eigenfunctions
$\Phi^S_{[1,1,1]}(\xi_1,\xi_{2},\xi_{3})$,
$\Phi^S_{[0,0,4]}(\xi_1,\xi_{2},\xi_{3})$  and A-eigenfunction
$\Phi^A_{[0,2,4]}(\xi_1,\xi_{2},\xi_{3})$, at $A=4$ (left, middle,
and right panels, respectively ). Some maxima and minima positions
of these functions are connected by black and gray lines and
duplicated in lower panels: two tetrahedrons forming a
\textit{stella octangula} for
$\Phi^S_{[1,1,1]}(\xi_1,\xi_{2},\xi_{3})$, a cube and an
octahedron for $\Phi^S_{[0,0,4]}(\xi_1,\xi_{2},\xi_{3})$, and a
polyhedron with 20 triangle faces (only 8 of them being
equilateral triangles) for
$\Phi^A_{[0,2,4]}(\xi_1,\xi_{2},\xi_{3})$. } \label{table5}
\end{figure}

The SCR algorithm was implemented in MAPLE 14 on Intel Core i5 CPU 660 3.33GHz, 4GB 64 bit,
to generate
first $11$ symmetric (antisymmetric) functions up to $\Delta E_j=12$ at $A=6$ with CPU time 10 seconds (600 seconds),
that together with a number of functions in dependence of number of particles given in Table \ref{10}
 demonstrates efficiency and complexity of the algorithm.

The examples of generated total symmetric and antisymmetric
$(A-1)$-dimensional oscillator functions are presented in Tables
\ref{18} and \ref{18f}. Note that  for $A=4$, the first four
states from Table \ref{18} are similar to those of the
translation-invariant model without excitation of the
center-of-mass variable \cite{Moshinsky2}.

As an example,  in Figs. \ref{table3} and \ref{table4} we show isolines of the first eight  S and A oscillator eigenfunctions $\Phi^S_{[i_1,i_2]}(\xi_1,\xi_{2})$ and $\Phi^A_{[i_1,i_2]}(\xi_1,\xi_{2})$
for $A=3$, calculated at the second step of the algorithm.
One can see that the S (A) oscillator eigenfunctions
are symmetric (antisymmetric) with respect to reflections from three straight lines.
The first line (labelled `23') corresponds to the permutation $(x_2,x_3)$ and is rotated by $\pi/4$ counterclockwise
with respect to the axis $\xi_1$.
The second and the third  lines (labelled `12' and `13') correspond to the permutations $(x_1,x_2)$ and $(x_1,x_3)$ and are rotated by $\pi/3$ clockwise
and counterclockwise with respect to the first line.
These lines divide the plane into six sectors,
while the symmetric (antisymmetric) oscillator eigenfunctions,
calculated at the first step of the algorithm,
which are symmetric (or antisymmetric) with respect to reflections from the first line, generate the division of the plane into two parts.
This illustrates the isomorphism between the symmetry group of
an equilateral triangle $D_3$ in $\textbf{R}^2$ and the 3-body permutation group $S_3$
(A = 3).

\begin{table}[t]
\caption{The degeneracy  multiplicities $p$  from (\ref{nbetab}),  $p_s=p_a$ and $p_S=p_A$
of s-, a-, S-, and A-eigenfunctions
of the oscillator energy levels $\Delta E_j =E_j^\bullet-E_1^\bullet$, $\bullet=\emptyset,s,a,S,A$.
}\label{10}
\begin{center}\begin{tabular}{||c|c|c||c|c|c||c|c|c||c|c|c||c||} \hline
 \multicolumn{3}{|c||}{A=3}  & \multicolumn{3}{|c||}{A=4}   & \multicolumn{3}{|c||}{A=5}   & \multicolumn{3}{|c||}{A=6}& $\Delta E_j$\\ \hline
$p$&$p_s,p_a$&$p_S,p_A$&$p$&$p_s,p_a$&$p_S,p_A$&$p$&$p_s,p_a$&$p_S,p_A$&$p$&$p_s,p_a$&$p_S,p_A$&\\  \hline
 1&1 &1  & 1 &1 & 1& 1  &1& 1& 1  & 1&1 &0   \\
 2&1 &0  & 3 &1 & 0& 4  &1& 0& 5  & 1&0 &2   \\
 3&2 &1  & 6 &2 & 1&10  &2& 1& 15 & 2&1 &4  \\
 4&2 &1  &10 &3 & 1&20  &3& 1& 35 & 3&1 &6  \\
 5&3 &1  &15 &4 & 2&35  &5& 2& 70 & 5&2 &8   \\
 6&3 &1  &21 &5 & 1&56  &6& 2&126 & 7&2 &10  \\
 7&4 &2  &28 &7 & 3&84  &9& 3&210 &10&4 &12  \\
\hline
\end{tabular}\end{center}\end{table}
Figure \ref{table5}  shows examples of profiles of S and A
oscillator eigenfunctions for $A=4$. Note that  four maxima
(black) and four minima (grey) of the S eigenfunction
$\Phi^S_{[1,1,1]}(\xi_1,\xi_{2},\xi_{3})$ are positioned at the
vertices of two tetrahedrons forming a \textit{stella octangula},
with the edges shown by black and grey lines, respectively. Eight
maxima and six outer minima for S eigenfunction
$\Phi^S_{[0,0,4]}(\xi_1,\xi_{2},\xi_{3})$ are positioned at the
vertices of a cube and an octahedron, the edges of which are shown
by black and grey lines, respectively. The positions of twelve
maxima of the A oscillator  eigenfunction,
$\Phi^A_{[0,2,4]}(\xi_1,\xi_{2},\xi_{3})$ coincide with the
vertices of a polyhedron with 20 triangle faces (only 8 of them
being equilateral triangles) and 30 edges, 6 of them having the
length 2.25 and the other having the length 2.66. The above shapes
of eigenfunctions  illustrate the isomorphism between the
tetrahedron group $T_d$ in ${\bf R}^{3}$ and the 4-particle
permutation group $S_4$  (A = 4), discussed in
\cite{KramerMoshinsky} in the case of $d=3$.

The degeneracy multiplicity (\ref{nbetab}), i.e., number $p$ of
all states  with the given energy $E_j$ of low part of spectra,
the numbers $p_{s}$ ($p_a$) of the states, symmetric
(antisymmetric) under permutations of $A-1$ relative coordinates
together with the total numbers  $p_{S}$ ($p_{A}$) of the states,
symmetric (antisymmetric) under permutations of $A$ initial
Cartesian coordinates are summarized in Table \ref{10}. Note that
the S and A states with $E'=E_1^{S,A}+2$ do not exist. The numbers
$p_{s}$ $(p_{a})$ are essentially smaller than the total number
$p$ of all states, which simplifies the procedure of constructing
S (A) states with possible excitation of the center-of-mass degree
of freedom and allows the use of a compact basis with the reduced
degeneracy $p_{S}$ ($p_{A}$) of the S (A) states in our final
calculations. For clarity, in the case $A=3$, $d=1$, the S(A)-type
functions generated by the \textit{SCR} algorithm, in polar
coordinates $\xi_1=\rho\cos\varphi$, $\xi_2=\rho\sin\varphi$ are
expressed as:
\begin{eqnarray*}&&
\Phi^{S(A)}_{k,m}(\rho,\varphi)=C_{km}(\rho^2)^{3m/2}\exp(-\rho^2/2)L_{k}^{3m}(\rho^2)\begin{array}c\cos\\ [-1mm] \sin \end{array}(3m(\varphi+\pi/12)),
\end{eqnarray*}
where $C_{km}$ is the normalization constant, $L_{k}^{3m}(\rho^2)$ are the Laguerre polynomials \cite{stigun},
$k=0,1,...$,  $m=0,1,...$ for S states, while $m=1,2,...$ for A states,
that are classified by irr of the $D_{3m}$-symmetry group.
The corresponding  energy levels
$E^{S(A)}_{k,m}=2(2k+3m+1)=E^{s(a)}_{[i_1,i_2]}=2(i_1+i_2+1)$ have
the degeneracy  multiplicity $K+1$, if the energy
$E^{S(A)}_{k,m}-E_1^{S(A)}=12K+K'$, where $K'=0,4,6,8,10,14$. For
example, in Figs. \ref{table3} and \ref{table4} we show the wave
functions $\Phi^S_{3,0}(\rho,\varphi)$ and
$\Phi^S_{0,2}(\rho,\varphi)$ (or $\Phi^A_{3,1}(\rho,\varphi)$ and
$\Phi^A_{0,3}(\rho,\varphi)$) labelled with 6 and 7, corresponding
to the energy levels $E^{S(A)}_{k,m}-E_1^{S(A)}=12$ with the
degeneracy $K=2$, while the functions labelled with $1,2,3,4,5,8$
are nondegenerate  ($K=1$). So, the eigenfunctions of  the
A-identical particle  system  in one dimension are degenerate in
accordance with ~\cite{LL68}, and this result disagrees with
nondegenerate ansatz solutions ~\cite{Wang}.

\section{Conclusion}
We considered a model of $A$ identical particles bound by the
oscillator-type potential under the influence of the external
field of a target in the new symmetrized coordinates. The
constructive SCR algorithm of symmetrizing or antisymmetrizing the
$A-1$-dimensional harmonic oscillator basis functions with respect
to permutations of A identical particles was described.
One can see that the transformations of $(A-1)$-dimensional oscillator basis functions from the symmetrized coordinates to the Jacobi coordinates, reducible to permutations of coordinates and $(A-1)$-dimensional finite rotation (\ref{b}), are implemented by means of the $(A-1)$-dimensional oscillator Wigner functions \cite{PST1981}.
Typical examples were analyzed, and a correspondence between the
representations of the symmetry  groups $D_3$ and $T_d$ for $A=3$
and $A=4$ shapes is displayed. It is shown that one can use the
presented SCR algorithm, implemented using the MAPLE computer
algebra system, to construct the basis functions in the closed
analytical form. However, for practical calculations of matrix
elements between the basis states, belonging to the lower part
of the spectrum, this is not necessary. 
The application of the developed approach and algorithm for solving the problem
of tunnelling clusters through barrier potentials of a target
is considered in our forthcoming paper~~\cite{Tunneling2013}.
The proposed approach can be adapted  to the
analysis of tetrahed\-ral-symmetric  nuclei,
quantum diffusion of molecules and
micro-clusters through surfaces,
and  the fragmentation in producing
neutron-rich light nuclei.

The authors thank Professor V.P. Gerdt  for collaboration.
The work was supported partially by grants 13-602-02 JINR, 11-01-00523 and 13-01-00668  RFBR
and the Bogoliubov-Infeld program.

\end{document}